\documentclass[twocolumn,floatfix,prd,showpacs,amsmath,nofootinbib,preprintnumbers]
{revtex4}

\usepackage{graphicx}

\newcommand{\vz}{\vec{\zeta}}
\newcommand{\mpch}{\mbox{Mpc}/h}
\newcommand{\impch}{h/\mbox{Mpc}}
\newcommand{\Deg}{^{\circ}}
\newcommand{\be}{\begin{equation}}
\newcommand{\ee}{\end{equation}}
\newcommand{\bea}{\begin{eqnarray}}
\newcommand{\eea}{\end{eqnarray}}
\newcommand{\vs}{\nonumber\\}

\begin{document}

\title{Universal Weak Lensing Distortion of Cosmological Correlation Functions}

\author{Scott Dodelson}
\email{dodelson@canis.fnal.gov}
\affiliation{Center for Particle Astrophysics, Fermi National Accelerator Laboratory,
Batavia, IL~~60510-0500, USA,\\
Department of Astronomy \& Astrophysics,~The University of Chicago, Chicago, IL~~60637-1433, USA,}

\author{Fabian Schmidt}\email{fabians@uchicago.edu}
\affiliation{Department of Astronomy \& Astrophysics,~The University of Chicago, Chicago, IL~~60637-1433,\\
Kavli Institute for Cosmological Physics, Chicago, IL~~60637-1433}

\author{Alberto Vallinotto}\email{vallinot@iap.fr}
\affiliation{Institut d'Astrophysique de Paris, UMR7095 CNRS,
Universit\'{e} Pierre et Marie Curie, 98 bis Boulevard Arago, 75014
Paris.}

\date{\today}

\begin{abstract}
Gravitational lensing affects observed cosmological correlation functions because observed images do not coincide with true source locations. We treat this universal effect in a general way
here, deriving a single formula that can be used to determine to 
what extent this effect distorts any correlation function. We then apply the general formula to the correlation functions of galaxies, of the 21-cm radiation field, and of the CMB. 
\end{abstract}

\maketitle

\section{Introduction}

Gravitational lensing has emerged as a powerful cosmological tool. The spectacular images provided by strong
lensing~\cite{1996ApJ...471..643K,Kneib:2001qy,Bradac:2006er} together with the robust parameter determination anticipated with weak lensing~\cite{Kaiser:1991qi,Mellier:1998pk,Bartelmann:1999yn,Hu:1999ek,Huterer:2001yu,Abazajian:2002ck,Takada:2003ef,Refregier:2003xe} are just two of
the exciting developments in the field. When astronomers look for lensing, they have demonstrated that they can
find it. Lensing can also contaminate observations of correlation functions because the images of objects are displaced from the true source positions. For example, it has long been known that lensing
smoothes out the spectrum of the cosmic microwave background (CMB)~\cite{Seljak:1995ve,Lewis:2006fu,Smith:2007rg}. The effect of lensing on galaxy-galaxy correlations 
has been re-examined by several groups recently~\cite{Moessner:1997vm,Matsubara2000,Hui:2007cu,Vallinotto:2007mf,LoVerde:2007ke,Hui:2007tm,Schmidt:2008mb}. And a number of groups are anticipating the importance of lensing on future observations of neutral hydrogen via long wavelength measurements of the redshifted 21 cm line~\cite{Mandel:2005xh,Zhang:2006fc,Lewis:2007kz}.

In this work we present a unified treatment of the lensing contamination of correlation functions. First, we show that weak lensing contaminates \textit{any} cosmological correlation function. Whereas previous studies have quantified its effect on some specific correlation functions (for instance, CMB temperature and polarization \cite{Lewis:2006fu} and baryon oscillations through the galaxy correlation function \cite{Vallinotto:2007mf}), we focus here on the \textit{generality} of the result. This general framework allows us to understand when and for what correlation functions weak lensing represents a sizeable or a negligible effect. By working in real space, we also hope to provide physical intuition for some of the properties of the lensing effects, which is harder to obtain from a pure $k$-space calculation.
Second, we consider the dependence of the lensing effect on the orientation
of the source separation with respect to the line of sight. We also include the 
lensing-induced time delay as the 
longitudinal companion effect to the transverse lensing deflections.
Finally, we show how the general framework presented here can be applied to different correlation functions: we retrieve previous results for the CMB, supplement and correct previous galaxy-galaxy results~\cite{Vallinotto:2007mf}, and apply the formalism to the anisotropy induced by lensing of 21-cm radiation. 

The paper is organized as follows. In Sec.~\ref{Sec:GenRes3D} a general formalism to calculate the weak lensing contribution to cosmological correlations is introduced along with a more physical understanding of the different quantities involved. In Sec.~\ref{Sec:Apps} the general formalism developed is applied to three different cases: correlation of galaxies, of the cosmic microwave background and of the 21-cm radiation. Finally, Sec.~\ref{Sec:Conclusions} contains some discussion together with concluding remarks. The calculations are relegated to the appendices.

\section{General Formalism}
\label{Sec:GenRes3D}

Weak lensing, i.e. the effect of small potential differences in the intervening space 
on the path of light, consists to first order of two effects: a {\it transverse
deflection} of the photons from a straight path, and a {\it time delay} (Shapiro
delay) along the path. Together, these effects result in a three-dimensional displacement of the apparent source position.

Consider two physical observables $A(\vec{x}_a)$ and $B(\vec{x}_b)$ and denote the \textit{observed} values of $A$ and $B$ as $\tilde{A}(\vec{x}'_a)$ and $\tilde{B}(\vec{x}'_b)$. In this section we keep the nature of these cosmological observables \textit{completely unspecified}. The form of the results depends neither on the actual physical observables that are being correlated nor on the nature of the source of the signal that is used to measure them.
Since the geodesics light travels on are perturbed by the intervening distribution of
matter, the measured values of the observables refer to physical points that 
are displaced with respect to the observed ones. 

\begin{figure}[t]
\includegraphics[width=0.5\columnwidth]{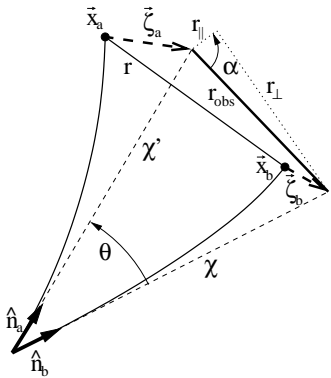}
\caption{Gravitational lensing bends photon trajectories. The sources, physically located at $\vec{x}_a$ and $\vec{x}_b$ are observed at $\vec{x}_a+\vec{\zeta}_a$ and $\vec{x}_b+\vec{\zeta}_b$ (in direction $\hat{n}_a$ and $\hat{n}_b$). $\alpha$ denotes the angle with the line of sight of the observed separation vector, $\vec{r}_{obs}$. \label{fig:deflection_geometry}}
\end{figure}

Consider the arrangement depicted in Fig.~\ref{fig:deflection_geometry}. Since 
gravitational lensing perturbs the photons' trajectories, the two sources -- 
located at $\vec{x}_a$ and $\vec{x}_b$ -- are observed at 
$\vec{x}'_a=\vec{x}_a+\vec{\zeta}_a$ and $\vec{x}'_b=\vec{x}_b+\vec{\zeta}_b$. 
The observed distance 
$\vec{r}_{\textrm{obs}}$ between the sources will in general differ from the 
true distance $\vec{r}$. Whereas isotropy demands that the unlensed correlation
function depends only on the distance $r$, the lensed correlation function 
depends on the angle $\alpha$ that $\vec{r}$ makes with the line of sight. 
Note that any difference in the line-of-sight angle of $\vec{r}$ and 
$\vec{r}_{obs}$ is a higher order correction, and we hence consider them to be
equal for the perturbative approach adopted in this paper.
Whenever the correlation function of two cosmological observables is measured, 
such measurement is subject to a modification arising because of the 
displacement of the apparent source positions $\vec{x}'_a$, $\vec{x}'_b$:
\begin{equation}\label{GenRes:xcorr1}
\langle \tilde{A}(\vec{x}'_a) \tilde{B}(\vec{x}'_b)\rangle=\langle A(\vec{x}'_a-\vec{\zeta}_a) B(\vec{x}'_b-\vec{\zeta}_b)\rangle.
\end{equation}
By Fourier transforming Eq.~(\ref{GenRes:xcorr1}) and introducing the power spectrum of the two observables 
\begin{equation}\label{GenRes:PS}
\langle A(\vec{k}_a,\eta_a) B^*(\vec{k}_b,\eta_b) \rangle\equiv(2\pi)^3\delta(\vec{k}_a-\vec{k}_b)P_{AB}(k_a, \eta_a, \eta_b),
\end{equation}
it is straightforward to show that
\begin{eqnarray}
\langle\tilde{A}(\vec{x}_a)\tilde{B}(\vec{x}_b)\rangle&=&
\int\frac{d^3k}{(2\pi)^3}
e^{i\vec{k}\cdot(\vec{x}_a-\vec{x}_b)}P_{AB}(k, \eta_a, \eta_b)\nonumber\\
&\times&
\langle e^{i\vec{k}\cdot(\vec{\zeta}_a-\vec{\zeta}_b)}\rangle,
\label{GenRes:expcorrfunc}
\end{eqnarray}
assuming that the observables $A$, $B$ are uncorrelated with the lensing 
deflection field $\vec{\zeta}$. Here $\eta_A$ and $\eta_B$ denote the conformal
times at which light was emitted at the sources to reach us today.

Equation (\ref{GenRes:expcorrfunc}) allows us to express the effect that weak lensing has on the unlensed correlation function. In other words, it quantifies the effect that the matter distribution -- responsible for lensing -- has on the observed distribution of the physical observables. 
Note that the weak lensing modification depends only on the difference $\vz_a-\vz_b$ 
of the deflections in both observables, not on the absolute magnitude of each. 
Assuming a flat Universe, as we do throughout, these distortions are given
by the following integrals along the line of sight:
\begin{eqnarray}
 \zeta_{a,\perp}^j&=&2\int_0^{\chi_a} d\chi(\chi_a-\chi)\nabla^j\Phi\left[\chi\vec\theta_a,\chi\right]\label{eq:perp}\\
\zeta_{a,\parallel}&=&-2\int_0^{\chi_a}d\chi\,\Phi\left[\chi\vec\theta_a,\chi\right] \label{eq:par}.
\end{eqnarray} 
Here the subscripts $\perp$ and $\parallel$ indicate directions transverse 
and parallel to the line of sight, respectively. 
In the transverse case, there are two such direction, indexed here with $j=1,2$. The subscript $a$ refers to the position of the observable $A$, fully specified by the direction $\vec{\theta}_a$ and the comoving distance $\chi_a$. In the integral over $\chi$, the potential is to be evaluated along the line of sight so its argument is $\vec x(\chi)=[\chi\vec\theta_a,\chi]$ valid in the small angle limit we consider here. There is one important difference between the parallel distortion $\zeta_\parallel$ and the transverse distortion $\zeta_\perp$: the latter has an extra factor of $(\chi_a-\chi)\nabla$. The potential typically varies on small scales, so this factor is of order $\chi/r \gg 1$. The transverse distortion therefore produces the dominant change in correlation functions.

By neglecting all non-Gaussianity present in the lensing field, it is possible 
to evaluate Eq.~(\ref{GenRes:expcorrfunc}) to
all orders, thus generalizing the results obtained by Challinor and Lewis 
\cite{Lewis:2006fu} in the case of CMB lensing. 
However, it is not clear to what extent this improves the
accuracy, since non-Gaussian terms, induced by gravity during structure 
formation, appear in the expansion at all orders above
the second and may play a significant role especially at low redshift and 
small scales. This issue can be resolved by tracing light rays through
N-body simulations of cosmological volumes (e.g., \cite{CarboneEtal}). 
In the case of CMB lensing, there are indications that deviations from the 
predictions of the calculation of \cite{Lewis:2006fu} which neglect 
non-Gaussianities in the deflection field begin to appear at
$\ell \gtrsim 2000$ in the deflection field.
A detailed analysis of such a general treatment and of its 
range of validity, requiring ray-tracing measurement of the probability 
density function (PDF) of the displacements $\vec{\zeta}$, will be the focus 
of a forthcoming work. We therefore briefly outline this general formalism 
in appendix \ref{App:Exact} and for the rest of this work we will follow the
perturbative approach. 

In order to proceed, we expand the exponential in Eq.~(\ref{GenRes:expcorrfunc})
to second order in the deflection. Note that all odd terms in this expansion
vanish due to symmetry, so that the expansion is good to fourth order. This is 
sufficient to quantify the weak lensing impact on the correlation function 
for sufficiently large separations.
Expanding Eq.~(\ref{GenRes:expcorrfunc}) above we then have
\begin{eqnarray}
\langle\tilde{A}\tilde{B}\rangle
&\approx &\langle A B\rangle 
+ \frac{Z^{ij}}{r^2}
\nonumber\\ 
&\times&\left[ r_i r_j \frac{d^2 \langle AB\rangle}{dr^2}+ r\frac{d \langle AB\rangle}{dr}\left(\delta_{ij}-\frac{r_i r_j}{r^2}\right)\right],\label{GenRes:Corrfunctexpansion}
\end{eqnarray}
where $\vec{r}=\vec{x}_a-\vec{x}_b$ denotes the observed separation between the sources and the $3\times3$ distortion correlation tensor is defined as
\begin{eqnarray}
Z^{ij} &\equiv& \frac{\langle \zeta_a^i\zeta_a^j \rangle + \langle \zeta_b^i\zeta_b^j \rangle}{2}-\langle\zeta_a^i\zeta_b^j\rangle\nonumber\\
&\equiv&
\left(\begin{matrix}
 T-D/2 & 0 & -V_a\cr
0 & T+D/2 & 0 \cr
-V_b & 0 & S 
\end{matrix}\right).
\label{Zdef}
\end{eqnarray}
The second definition here of the matrix elements holds if we choose $\vec x_a$ to lie along the $\hat z$ axis and both positions to lie in the $x-z$ plane, so that $\vec{x}_a$ and $\vec{x}_b$ have coordinates $\vec{x}_a=(0,0,\chi_a)$ and 
$\vec{x}_b=(r_{\perp},0,\chi_a+r_{\parallel})$ respectively. Note that 
due to the azimuthal symmetry of the lensing 
displacements around the line of sight, we can always choose such coordinates without loss of generality.
In appendix \ref{App1} we derive expressions for the elements of the distortion tensor using the Limber approximation. The correlation of the distortions transverse to the line of sight decompose into a piece which is similar along both $x$ and $y$ directions
\begin{equation}
T(\chi,r_\parallel,r_\perp) = T_1(\chi,r_\parallel) + T_2(\chi,r_\perp) 
\end{equation}
with
\begin{eqnarray}
T_1(\chi,r_\parallel) &\equiv& r_\parallel^2 \int_0^{\chi} d\chi' \int_0^\infty \frac{dk k^3P_\Phi(k,\chi')}{2\pi} \label{eq:t1}
\\
T_2(\chi,r_\perp)&\equiv &\int_0^{\chi} d\chi' \int_0^\infty \frac{dk k^3P_\Phi(k,\chi')}{\pi} (\chi-\chi')^2 
\vs &&\times
\left(1-J_0(kr_\perp\chi'/\chi)\right)\label{eq:t2}
\end{eqnarray}
and one which differs in the two transverse directions:
\begin{eqnarray}
D(\chi,r_\perp) &=& 2\int_0^{\chi} d\chi' (\chi-\chi')^2 \vs &&\times\int_0^\infty \frac{dk k^3P_\Phi(k,\chi')}{\pi} 
J_2(kr_\perp\chi'/\chi).\label{eq:d}
\end{eqnarray}
Note that both $T_2$ and $D$, which depend only on the transverse distance between the two background objects, vanish in the limit that $r_\perp=0$. Since we will see that these two terms dominate the distortion of the correlation function, changes to the correlation functions have a characteristic dependence on the
\textit{projected} separation $r_\perp$, peaking when $r_\perp=r$.
The variance of the displacements along the line of sight due to time delay is:
\begin{eqnarray}
S(\chi,r_\perp) &=& 2\int_0^\chi d\chi' \int_0^\infty \frac{dk\,k P(k,\chi')}{\pi}\vs &&\times\left[1-J_0(kr_\perp \chi'/\chi)\right].\label{eq:s}
\end{eqnarray}
Finally, the displacement along the line of sight is slightly correlated with the transverse distortion; the relevant combination is 
\begin{equation}
V_a+V_b=2r_\parallel \int_0^\chi d\chi' \int_0^\infty \frac{dk\,k^2 P(k,\chi')}{\pi} J_1(kr_\perp \chi'/\chi).
\end{equation}

It is useful to estimate the order of magnitude of the corrections. First 
consider Eqs.~(\ref{eq:perp}, \ref{eq:par}). The RMS of the perpendicular 
distortion is of order $\Phi_{\rm RMS} \chi^2/r$, where $\chi$ is a 
typical cosmological distance, of order a Gpc, and $r$ is a typical distance 
over which the potential varies, typically much smaller. The RMS of the 
parallel distortion is seen to be smaller by a factor of $(r/\chi)$. These 
estimates translate well into the respective variances as encoded in the 
functions $T_2$ and $D$. Both $T_2$ and $D$ are seen from Eqs.~(\ref{eq:t2}, 
\ref{eq:d}) to be of order $k^3P_\Phi(k) k\chi^3$ since the line of sight 
variable $\chi'$ is order $\chi$.  The variance of the gravitational potential 
is roughly $\Delta^2_\Phi(k)\sim k^3P_\Phi(k)$, and the $k$ integral picks 
out values of $k\sim r^{-1}$. So both $T_2$ and $D$ are of order 
$\Delta^2_\Phi(1/r) (\chi^2/r)^2\times (r/\chi)$, where $\Delta^2_\Phi(k)$ 
is equal to $\Phi_{\rm RMS}^2 \sim 10^{-9}$ on large scales 
($k\lesssim 0.01\impch$) and suppressed on smaller scales due to the transfer 
function. The extra factor of $r/\chi$ 
is the standard Limber suppression due to cancellation along the line of 
sight (only modes with $k_z$ small contribute to the variance).  
On large scales, the relevant dimensionless quantities $T_2/r^2$, $D/r^2$ 
thus are of order $\Phi_{\rm RMS}^2 \chi^3/r^3 \sim 10^{-6}\, (100\:\mpch/r)^3$.
They increase towards small scales, however not as quickly as $r^{-3}$, since
the variance of the potential is suppressed by the transfer 
function towards smaller scales. In other words, due to the suppressed power 
on small scales,
photons from both directions are deflected more and more coherently, and 
$r_{obs}$ is very close to $r$ which is reflected by reduced values
of $T_2$ and $D$.

The first term in the transverse deflection correlation, 
$T_1$ as defined in Eq.~(\ref{eq:t1}), represents the difference between the 
lensing experienced by A and B due to their different distances from us, so 
this part of the transverse lensing is suppressed by a factor of 
$r_\parallel^2/\chi^2$ and can usually be neglected. Similar estimates show 
that $S$ and $V_a+V_b$ are both smaller than the transverse variances $T_2$ 
and $D$ by a factor of $(r/\chi)^2$ as expected from a cursory examination 
of Eqs.~(\ref{eq:perp},\ref{eq:par}). So we expect, and numerical work 
confirms this, that $T_2$ and $D$ dominate the corrections to correlation 
functions, and the corrections will be of order 
$\Delta^2_\Phi(1/r) (\chi/r)^3$. 

\begin{figure}
\includegraphics[width=0.49\textwidth]{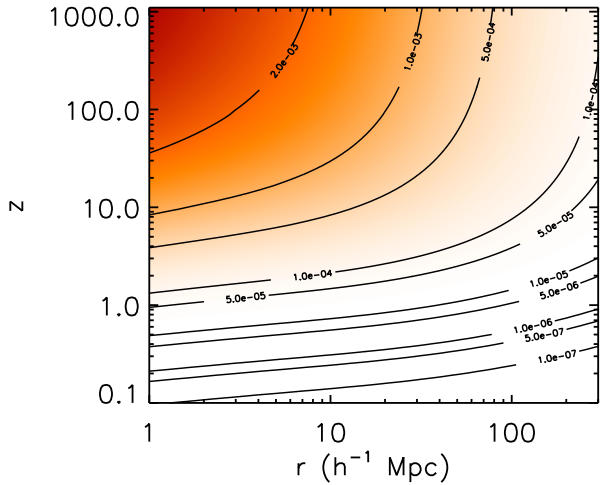}
\caption{Contour plot of $T_2/r^2$ [Eq.~(\ref{eq:t2})] determining the relative transverse
displacement as a function of the source separation $r$ and source redshift $z$ 
(assumed
equal for both sources, so $r_\perp=r$).  }\label{fig:Ta}
\end{figure}

\begin{figure}
\includegraphics[width=0.49\textwidth]{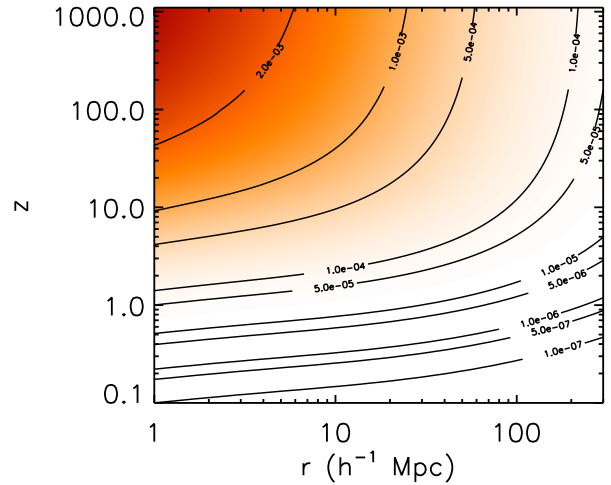}
\caption{Same as Fig.~\ref{fig:Ta} but for $D/r^2$, where $D$ is defined in Eq.~(\ref{eq:d}). }\label{fig:d}
\end{figure}

Therefore, corrections to cosmological correlations due to weak lensing are small and given by:
\begin{eqnarray}
\langle\tilde{A}\tilde{B}\rangle(r,r_\perp,\chi)
\approx
\langle A B\rangle(r,\chi)  
+ T_2(\chi,r_\perp)\frac{2}{r}\frac{d\langle AB\rangle}{dr}
 \vs 
+\left(T_2(\chi,r_\perp)-\frac{D(\chi,r_\perp)}{2}\right) \frac{r_\perp^2}{r^2} r \frac{d}{dr}\left[
\frac{1}{r}\frac{d\langle AB\rangle}{dr}\right].\vs
\label{GenRes:LH}
\end{eqnarray}
Two general features of this equation are worth pointing out: first, if the
correlation function is close to a pure power-law, both derivative terms
will be of order $\langle AB\rangle/r^2$. Hence, the relative 
lensing-induced effect on the correlation function will be given by
$T_2/r^2$ and $D/r^2$, which are shown in Fig.~\ref{fig:Ta} and \ref{fig:d}. 
Second, in case of an oscillating correlation function, the lensing
contribution can be amplified by large values of $d\langle AB\rangle/dr$
and $d^2\langle AB\rangle/dr^2$. Further,
lensing will tend to smooth out the oscillations: at a local minimum of $\langle AB\rangle$,
the first derivative vanishes, while the second derivative is positive,
so that the observed correlation is increased by lensing. At a local maximum, 
the opposite holds.
This feature is already well-known in $\ell$-space in case of the CMB.

Equation (\ref{GenRes:LH}) holds for any cosmological correlation function. It applies equally well to point-like sources and to diffuse backgrounds, to galaxies and QSO, to CMB and to the 21-cm radiation (albeit in the case of discrete sources magnification bias effects might provide the dominant distortion of the correlation function). Given a particular matter distribution, it is sufficient to evaluate the functions $T_2$ and $D$ \textit{once} to be able to calculate the effect that weak lensing has on the correlation function of \textit{any} cosmological observable. 

The functions $T_2/r^2$ and $D/r^2$ are shown for a concordance $\Lambda$CDM cosmology with $\Omega_m=0.3$ and $\Omega_{\Lambda}=0.7$ (which will be assumed throughout) in Figs.~\ref{fig:Ta} and \ref{fig:d}. We used the non-linear
matter power spectrum from \texttt{halofit} \cite{halofit} in the calculation.
As the positions $A$ and $B$ move out to higher redshift, the lensing effects become larger since the longer path lengths lead to larger RMS deflections. At 
fixed redshift, both $T_2$ and $D$ become larger as $r$ gets bigger since the 
displacements of the path from the two points to us cease to be correlated and 
hence experience independent deflections. However, as discussed above, the
relative effect of lensing distortions will generically be of order $T_2/r^2$,
$D/r^2$. These quantities decrease as 
$r$ increases, as illustrated in Figs.~\ref{fig:Ta} and \ref{fig:d}. Hence, 
in the absence of features in the unlensed correlation function considered,
the effect of lensing is likely to be most important on small scales. Since 
the perturbative expansion treatment we are using is valid under the condition 
that $(T_2/r^2,D/r^2) \ll 1$,  Figs.~\ref{fig:Ta} and \ref{fig:d} illustrate 
that the approximation used here is always applicable: it will yield a good
approximation to the size of the lensing effect. If the effect is only
marginally observable, this approximation should be sufficient. If however
the desired precision on lensing effect itself is high, as, e.g. in the case
of CMB lensing in order to improve cosmological parameter constraints, 
higher-order corrections will have to be taken into account.

\begin{figure}
\includegraphics[width=0.49\textwidth]{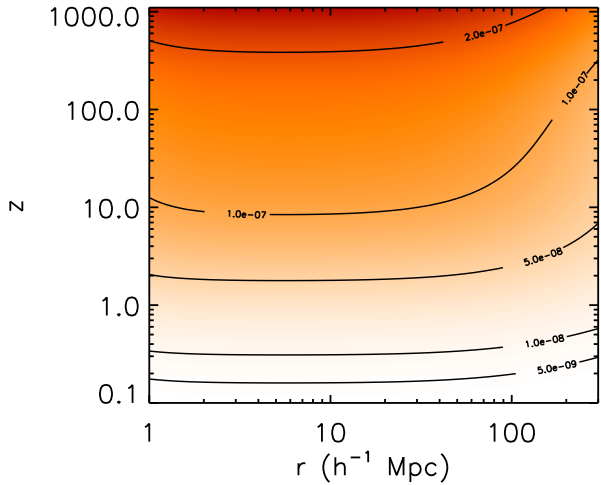}
\caption{Contour plot of $S/r^2$ [Eq.~(\ref{eq:s})] determining the relative longitudinal
displacement as a function of the source separation $r$ and source redshift $z$ 
(assumed
equal for both sources). }\label{fig:Sab}
\end{figure}

The value of the correlator of the longitudinal displacements $S/r^2 = Z^{33}/r^2$
is shown in Fig.~\ref{fig:Sab}. The longitudinal displacement effect (time delay) is clearly much smaller than the perpendicular one, as found earlier in \cite{CoorayHu}.

\section{Applications}\label{Sec:Apps}

We are now in a position to apply the above results to the correlation function of three different kinds of sources: galaxies, the CMB, and the 21-cm radiation background. The CMB is an angular measurement so all photons come from the same comoving distance meaning that $r_\parallel=0$ and the dependence is solely on $r=r_\perp$. For galaxies and 21cm, the orientation of the radius vector with respect to the line of sight is a free parameter, so the smoothing depends on 
$r_\perp$ even when $r$ is fixed.

\subsection{Galaxies}

\begin{figure}[ht]
\includegraphics[width=0.49\textwidth]{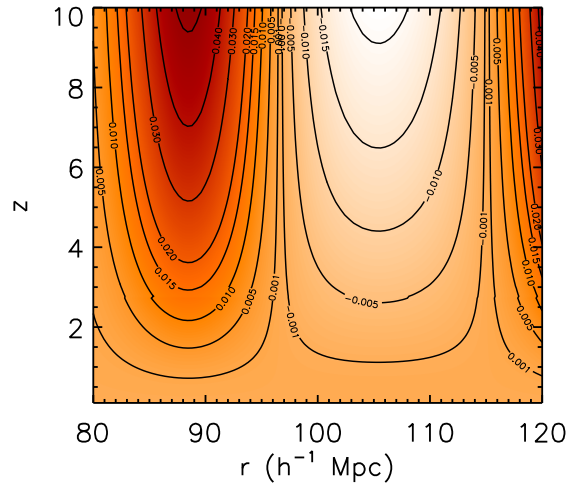}
\caption{Relative deviation from the unlensed correlation function of the 
galaxy correlation function including effects of lensing deflections,
near the baryon acoustic scale. Lensing smoothes the BAO feature in the 
galaxy correlation function at the percent level.} \label{fig:NewBao}
\end{figure}

The effect of weak lensing on the galaxy correlation function has been taken into account in previous work in the context of the analysis of the impact that lensing has on the determination of the sound horizon of baryon oscillations \cite{Vallinotto:2007mf}. The ratio of the lensing induced term to the unlensed correlation function is plotted in Fig.~\ref{fig:NewBao}. Lensing smoothes the BAO bump at the percent level.
Note that since the lensing effect multiplies derivatives of the unlensed
correlation function, the relative effect of lensing will be independent of
any galaxy bias, in contrast to the magnification bias, whose relative effect
is $\propto 1/b$ or $\propto 1/b^2$, depending on redshift \cite{Hui:2007cu}.
We point out that Eqs.~(6-8) of \cite{Vallinotto:2007mf} are recovered using Eq.~(\ref{GenRes:LH}) above once the dependence of the kernels on the angle $\theta$ is correctly taken into account in \cite{Vallinotto:2007mf}. 

\begin{figure}[ht]
\includegraphics[width=0.49\textwidth]{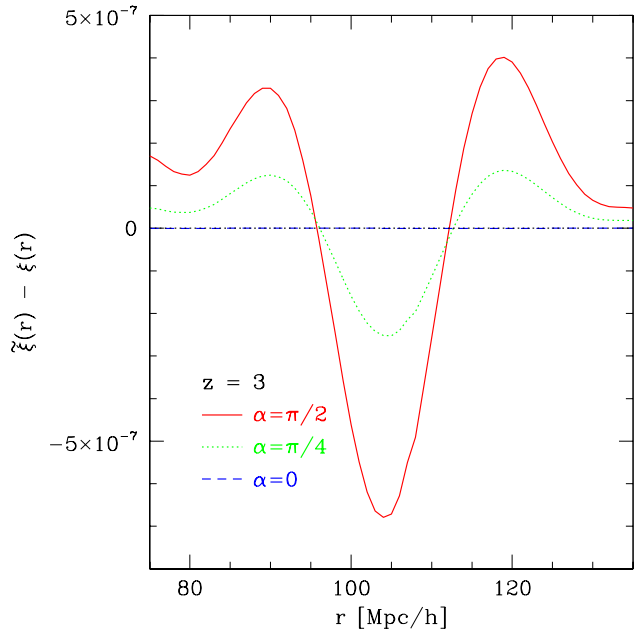}
\caption{Difference between the lensed and unlensed matter correlation 
function 
at $z=3$ near the BAO feature. The three lines shown are for different 
orientations of the radius vector with respect to the line of sight: perpendicular (solid/red), at 45$\Deg$ (dotted/green), and parallel to the line of sight
(dashed/blue). The lensing contribution shows a strong dependence on the
orientation of the separation vector.} \label{fig:deltaxi}
\end{figure}

Fig.~\ref{fig:deltaxi} shows the angular dependence of the correlation function for galaxies at redshift 3. The characteristic angular dependence illustrated in Fig.~\ref{fig:deltaxi} may make this effect detectable. Note that due to
the smallness of the time-delay effect ($S/r^2$, Fig.~\ref{fig:Sab}), the
lensing contribution essentially vanishes for $\alpha\rightarrow0$.

\subsection{Cosmic Microwave Background}

The smoothing effect of lensing on the CMB power spectrum was computed long ago~\cite{Seljak:1995ve} in multipole space. The formalism established here allows for a simple calculation of this same effect in angular space. 

The angular temperature correlation function of the CMB
is given by:
\be
w_{TT}(\theta) = \sum_{l=2}^{\infty} \frac{2l+1}{4\pi}
C^{TT}(l)\:P_l(\cos\theta),
\label{eq:wTT}
\ee
where $P_l$ denote the Legendre polynomials.
Applying the approach presented here, we can calculate the
lensing effect on the angular correlation function by
evaluating Eq.~(\ref{GenRes:LH}). Then, $\langle AB \rangle(r)$ stands for the
temperature correlation function $\xi_{TT}(r) = w_{TT}(r/\chi_{*})$, where
$\chi_{*}$ is the distance to the last scattering surface.
For this, we need the first and second derivatives of  $\xi_{TT}(r)$ which
are given by:
\bea
\frac{d}{dr} \xi_{TT}(r) &=& \frac{-\sin\theta}{\chi_{*}} \frac{d
w_{TT}}{d\cos\theta} \nonumber \\
\frac{d^2}{dr^2} \xi_{TT}(r) &=& \frac{-\cos\theta}{\chi_{*}^2} \frac{d
w_{TT}}{d\cos\theta} + \frac{\sin^2\theta}{\chi_{*}^2} \frac{d^2
w_{TT}}{d\cos\theta^2}
\label{eq:xiTT}
\eea
The derivatives of $w_{TT}$ with respect to $\cos\theta$ can be carried out using
the Legendre polynomial relation:
\be
P_l'(x) = \frac{l}{x^2-1} ( x P_{l}(x) - P_{l-1}(x) )
\ee

\begin{figure}[ht]
\includegraphics[width=0.49\textwidth]{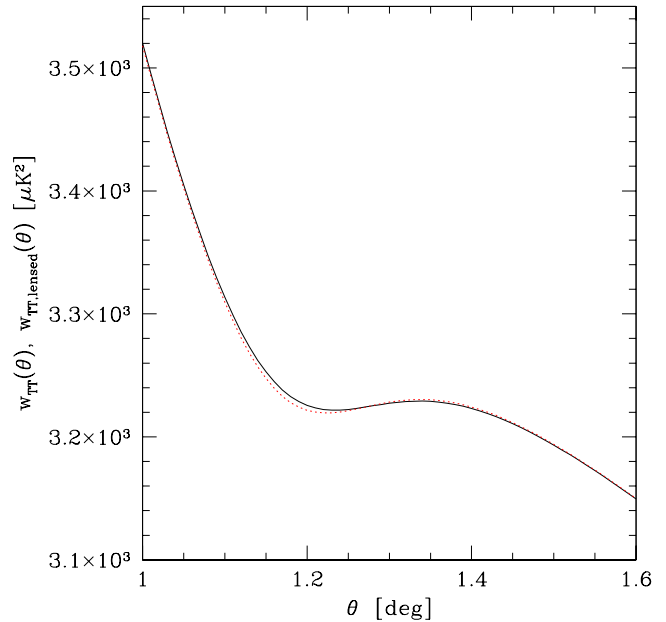}
\caption{Lensed (solid) and unlensed (dotted) CMB angular correlation functions.A slight smoothing effect washing out the baryon acoustic feature can be
seen. \label{fig:CMBlensing1}}
\end{figure}

\begin{figure}[ht]
\includegraphics[width=0.49\textwidth]{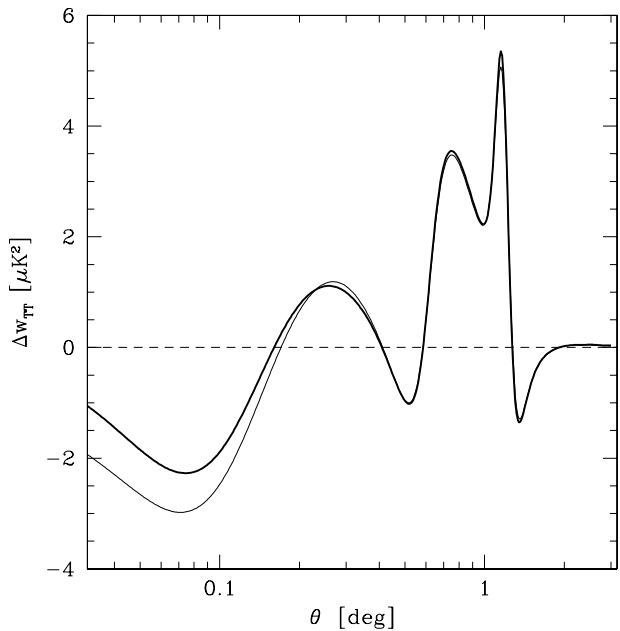}
\caption{Difference between the lensed and unlensed CMB angular correlation
function $w_{TT}(\theta)$. The thick line shows the calculation
using the first-order approach presented here, while the thin line is
the lensed $w_{TT}$ obtained from CAMB.\label{fig:CMBlensing2}}
\end{figure}

Fig.~\ref{fig:CMBlensing1} shows the lensed and unlensed CMB angular
correlation functions in the region of the baryon acoustic feature.
One can discern a slight smoothing effect of lensing in real space, as 
pointed out after Eq.~(\ref{GenRes:LH}).
Fig.~\ref{fig:CMBlensing2} shows the difference between the lensed and
unlensed correlation functions calculated in this approach (thick line). The
unlensed $C^{TT}(l)$ were obtained from CAMB \cite{Lewis:1999bs}. 
In the figure we also show the correlation function obtained from the 
lensed $C^{TT}(l)$ given by CAMB using Eq.~(\ref{eq:wTT}) (thin line). 
Clearly, the calculations in the two different approaches agree for $\theta \gtrsim 0.1\Deg$
corresponding to $r \gtrsim 15\mpch$. Note that results from N-body simulations
\cite{CarboneEtal} show deviations from the results of the CAMB code 
for $\ell \gtrsim 2000$, roughly corresponding to $\theta \lesssim 0.03\Deg$;
however, it is not straightforward to convert the scale where deviations
appear in multipole space to a corresponding angular scale in real space.

\subsection{21-cm Surveys}

In principle, redshifted 21 cm radiation encodes information about the 3D distribution of neutral hydrogen in the universe. This distribution is sensitive to both inhomogeneities in the matter and in the free electron density. It is not our purpose here to compute the complicated correlation function that results. Rather, we note that the real space calculation presented here is particularly simple to apply to any model of reionization (see Ref.~\cite{Mandel:2005xh} for a careful discussion of the complications that arise in Fourier space). 

\begin{figure}[t]
\includegraphics[width=0.49\textwidth]{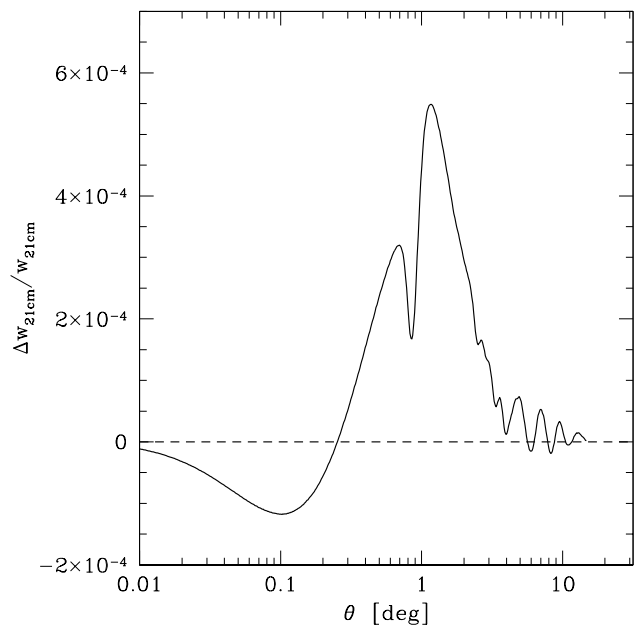}
\caption{Relative lensing effect on the angular correlation function
of 21cm emission from the dark ages. We use the cross-correlation coefficients
from \cite{Lewis21cm}. The redshift is $z=50$, and the bandwidth chosen
is 0.01~MHz.}
 \label{fig:21cm}
\end{figure}

For the purposes of this paper, we use the 21cm predictions for the ``dark ages''
before reionization calculated in \cite{Lewis21cm}. We use the 21cm spin temperature
correlation coefficients $C^{TT}(\ell)$ at $z=50$, in the same way as 
outlined above in the case of the CMB.
Fig.~\ref{fig:21cm} shows the relative lensing effect on the 21cm angular
correlation function. It is rather small compared to that of the CMB
due to the overall smoothness of the 21cm angular correlation function
in the dark ages. Note that at lower redshifts, $z\lesssim 12$, we expect 
the 21cm correlation function to show a stronger baryonic signature
\cite{Wyithe:2007rq}, and hence expect a higher lensing effect possibly of
importance to cosmological parameter constraints
\cite{McQuinn:2005hk,Bowman:2005hj,Santos:2006fp,Mao:2008ug}.

\section{Discussion}\label{Sec:Conclusions}

Gravitational lensing affects cosmological correlation functions in two ways. First, since the geodesics which photons travel on are sensitive to the distribution of matter between source and observer and since the latter is locally inhomogeneous thanks to structure formation, weak lensing acts by displacing the sources' observed (as opposed to true) positions. The correlation function that is measured from observed data therefore also includes the contribution from these lensing deflections. Second, weak lensing also acts on observations by focusing or defocusing the geodesics' congruences, thus altering the observed brightness of a source. This latter effect, known as \textit{magnification bias}, is generally larger than the former \cite{Matsubara2000,Hui:2007cu} but it applies only to pointlike sources (see e.g. \cite{Hui:2007cu, Vallinotto:2007mf, Schmidt:2008mb} for studies 
of the effect of magnification bias on two- and three-point correlation 
functions). Both these effects are small but potentially relevant in the present age of precision cosmology. In this work we focused on the first of these effects and we derived a \textit{general} perturbative formula that can be used to quantify it. 

The results of the present work depend only on the following two assumptions: that photons travel on geodesics and that higher order corrections to the correlation function (arising from higher order correlators of the lensing induced displacements) can safely be neglected. The first of this assumption has two consequences of opposite nature. The positive consequence is that the result obtained in this work is general and applies to \textit{any} cosmological correlation function, regardless of the nature of the source and of the physical observables that are being measured and correlated. Moreover, the correction terms that appear in Eq.~(\ref{GenRes:LH}) and that quantify the effect of lensing depend only on the power spectrum of matter: given a specific cosmological model they need to be evaluated only once and they can then be applied to any correlation function. As such, they also represent a map that will tell whether weak lensing will have a relevant or an irrelevant role in the measurement of a given correlation function \textit{before} the actual measurement is carried out. The negative consequence, on the other hand, is that this lensing-induced distortions represent a theoretical systematic that will \textit{always} be present regardless of the precision of the instruments used to carry out the observation. In other words, even if ``perfect observations'' free of any systematics could be carried out, this lensing-induced noise would still creep into the correlation function that would be calculated using those data. This lensing-induced modification of the correlation may be avoided only by reconstructing a map of the lensing
potential (e.g., \cite{Hirata:2002jy}). Another avenue to a possible disentanglement of this lensing distortion from the observed correlation function -- at least for sources that are not confined to a fixed redshift -- could be to exploit the angular dependence of the effect as seen in Fig.~\ref{fig:deltaxi}. Furthermore, since the lensing modification depends on the derivatives of the correlation function, correlation functions that are rapidly oscillating will in general be most affected by lensing, i.e. the features will be somewhat smoothed. Conversely, in quantities that are intrinsically uncorrelated, no correlation will be induced by lensing deflections. 

The second assumption entering the derivation of Eq.~(\ref{GenRes:LH}) is that terms of third and higher order are discarded when the exponential of Eq.~(\ref{GenRes:expcorrfunc}) is expanded. As shown in Appendix \ref{App1}, this assumption corresponds to Taylor expanding the coordinate dependence of the physical observables and to retain only terms up to second order, which in turn corresponds to neglecting all contributions arising from the non-Gaussianity of the displacements' PDF. Despite the fact that under the same assumptions the general formalism outlined in Appendix \ref{App:Exact} can be used to calculate the lensing distortion arising from the sum of \textit{all} the terms appearing in the exponential of Eq.~(\ref{GenRes:expcorrfunc}), it is however unclear to what extent this might really represent an improvement. If on one hand the sum of all the terms appearing in the exponential may give important contributions on small scales, it is also true that on such small scales departures from gaussianity of the displacements' PDF -- induced by the non-linear growth of structure at low redshift -- could play a relevant role. The range of applicability of such a non-perturbative method and the gain in precision that it would allow on small scales are the focus of a current project.

\acknowledgments
AV is supported by Agence Nationale de la Recherche (ANR). AV thanks Carlo Schimd, Emiliano Sefusatti and Jean-Philippe Uzan for useful discussions and comments. SD 
is supported by the Fermi Research Alliance, LLC under Contract No. DE-AC02-07CH11359 with the US Department of Energy.
FS thanks Wayne Hu for helpful discussions, and is
supported by the Kavli Institute for Cosmological Physics at the University of Chicago through grants NSF PHY-0114422 and NSF PHY-0551142 and an endowment from the Kavli Foundation and its founder Fred Kavli.

\bibliography{DSV3a}

\appendix

\section{Derivation of Eqs.~(\ref{GenRes:Corrfunctexpansion}-\ref{GenRes:LH})}\label{App1}

\subsection{Perturbative Approach}

Starting again from Eq.~(\ref{GenRes:expcorrfunc}) it is possible to expand the exponential keeping only terms up to second order 
\begin{eqnarray}
\langle\tilde{A}\tilde{B}\rangle
&\approx& \int\frac{d^3k}{(2\pi)^3}
e^{i\vec{k}\cdot(\vec{x}_a-\vec{x}_b)}P_{AB}(k, \eta_a, \eta_b)
\nonumber\\
&\times& \Big\langle 1+i\vec{k}\cdot(\vec{\zeta}_a-\vec{\zeta}_b)
\nonumber\\
&-&\frac{1}{2}[\vec{k}\cdot(\vec{\zeta}_a-\vec{\zeta}_b)][\vec{k}\cdot(\vec{\zeta}_a-\vec{\zeta}_b)] \Big\rangle,\label{App:correxpans2}
\end{eqnarray} 
The zero-th order (in $\vec{k}$) corresponds to the unlensed correlation function. Similarly, the first order term vanishes because $\langle\zeta\rangle=0$. Let's then move to consider the second term, which we'll denote as $\langle AB \rangle_2$. We can rewrite it as
\begin{eqnarray}
\langle AB \rangle_2&=&
-\frac{1}{2}\int\frac{d^3k}{(2\pi)^3}
e^{i\vec{k}\cdot(\vec{x}_a-\vec{x}_b)}P_{AB}(k, \eta_a, \eta_b)
\nonumber\\
&\times& k_i k_j \left[ \langle\zeta_a^i\zeta_a^j\rangle + \langle\zeta_b^i\zeta_b^j\rangle 
- \langle\zeta_a^i\zeta_b^j\rangle - \langle\zeta_b^i\zeta_a^j\rangle \right]\nonumber\\
&=&\left[\langle\zeta_a^i\zeta_b^j\rangle-\frac{\langle\zeta_a^i\zeta_a^j\rangle + \langle\zeta_b^i\zeta_b^j\rangle}{2}\right]\nonumber\\
&\times&\int\frac{d^3k}{(2\pi)^3}
e^{i\vec{k}\cdot(\vec{x}_a-\vec{x}_b)}P_{AB}(k, \eta_a, \eta_b)
k_i k_j
\end{eqnarray}
where we have used the fact that the displacement correlators do not depend on the integration variable $k$ to pull them out of the integral. The above integral can be rewritten as
\begin{eqnarray}
&&\int\frac{d^3k}{(2\pi)^3}
e^{i\vec{k}\cdot\vec{r}}P_{AB}(k, \eta_a, \eta_b)
k_i k_j\nonumber\\
&=&\int\frac{d^3k}{(2\pi)^3}\frac{1}{i^2}\frac{\partial^2}{\partial r_i \partial r_j}
e^{i\vec{k}\cdot\vec{r}}P_{AB}(k, \eta_a, \eta_b)\nonumber\\
&=&\frac{1}{i^2}\frac{\partial^2}{\partial r_i \partial r_j}\int\frac{d^3k}{(2\pi)^3}
e^{i\vec{k}\cdot\vec{r}}P_{AB}(k, \eta_a, \eta_b)\nonumber\\
&=&-\frac{\partial^2}{\partial r_i \partial r_j} \langle AB\rangle.
\end{eqnarray}
We can then notice that
\begin{eqnarray}
\frac{\partial^2 \langle AB\rangle}{\partial r_i \partial r_j}&=&
\frac{d^2 \langle AB\rangle}{dr^2} \frac{\partial r}{\partial r_i} \frac{\partial r}{\partial r_j} 
+\frac{d \langle AB\rangle}{dr}\frac{\partial^2 r}{\partial r_i \partial r_j}\nonumber\\
&=& \frac{r_i r_j}{r^2}\frac{d^2 \langle AB\rangle}{dr^2}\nonumber\\
&+&\frac{1}{r}\frac{d \langle AB\rangle}{dr}\left(\delta_{ij}-\frac{r_i r_j}{r^2}\right).
\end{eqnarray}
Putting all the pieces together we then get
\begin{eqnarray}
 \langle AB \rangle_2=\left[\frac{\langle\zeta_a^i\zeta_a^j\rangle + \langle\zeta_b^i\zeta_b^j\rangle}{2}-\langle\zeta_a^i\zeta_b^j\rangle\right]&&\nonumber\\
\times\Bigg[ \frac{r_i r_j}{r^2}\frac{d^2 \langle AB\rangle}{dr^2}\frac{1}{r}\frac{d \langle AB\rangle}{dr}\left(\delta_{ij}-\frac{r_i r_j}{r^2}\right)\Bigg]&&,
\end{eqnarray} 
which is exactly the result of Eq.~(\ref{GenRes:Corrfunctexpansion}) once we define the displacement correlator $Z^{ij}$ as
\begin{equation}
 Z^{ij}\equiv \frac{\langle\zeta_a^i\zeta_a^j\rangle + \langle\zeta_b^i\zeta_b^j\rangle}{2}-\langle\zeta_a^i\zeta_b^j\rangle.
\end{equation}  

Finally, let's notice that the same result can be obtained in a somewhat more intuitive way simply by Taylor expanding the coordinate dependence of the observables as
\begin{equation}
\tilde{A}(\vec{x}_a)
=A(\vec{x}_a+\vec{\zeta}_a)\approx A^a_0+\zeta_a^i A^a_{,i}+\frac{1}{2}\zeta_a^i\zeta_a^j A^a_{,ij},\label{GenRes:obsexp}
\end{equation}
where we use the shorthand notation $A^a_{,i}=\left. \partial A/\partial x^i \right|_{\vec{x}=\vec{x}_a}$,
and where the ``a'' index appearing on the displacement $\zeta_a^i$ and on the observables $A^a$ specifies that these quantities refer to the physical point $\vec{x}_a$.

\subsection{Decomposition of the Displacements' Correlator}

Consider a perturbed flat FRW spacetime with metric
\begin{equation}
 ds^2=a^2(\eta)\left[(-1-2\Psi)d\eta^2
+\delta_{ij}(1+2\Phi)dx^i dx^j \right].
\end{equation}
The lensing induced deflection of a source at distance $\chi_0$ is given by the following integrals along the (unperturbed) line of sight \cite{Dodelson,Bartelmann:1999yn}
\begin{eqnarray}
 \zeta_{a,\perp}^j&=&2\int_0^{\chi_0} d\chi(\chi_0-\chi)\nabla_{\perp}^j\Phi\nonumber\\
 &=&2i\int_0^{\chi_0} d\chi(\chi_0-\chi)
\int\frac{d^3k}{(2\pi)^3}k_{\perp}^j\Phi e^{i\vec{k}\cdot{\vec{x}(\chi)}},\quad\label{App:displacement_perp}\\
\zeta_{a,\parallel}&=&-2\int_0^{\chi_0}d\chi\,\Phi \nonumber\\
&=& -2\int_0^{\chi_0}d\chi \int\frac{d^3k}{(2\pi)^3}\Phi e^{i\vec{k}\cdot{\vec{x}(\chi)}}.\label{App:displacement_par}
\end{eqnarray} 
With the help of the Limber approximation, it is straightforward to show that the correlators are given by
\begin{eqnarray}
\langle \zeta_{a,\parallel} \zeta_{b,\parallel} \rangle&=&
4\int_0^{\textrm{min}\{\chi_{a}^0,\chi_{b}^0 \}}d\chi\nonumber\\
&\times&\int\frac{dk}{2\pi}k\,J_0(k\chi\theta)P_{\Phi}(k,\chi),\label{App:scalar}\\
\langle \zeta_{a,\perp}^j\zeta_{b,\parallel}\rangle&=&
-4i\int_0^{\textrm{min}\{\chi_{a}^0,\chi_{b}^0 \}} d\chi \, (\chi_{a}^0-\chi)
\nonumber\\
&\times&\int\frac{d^2\vec{k}_{\perp}}{(2\pi)^2}\, k_{\perp}^j \, e^{i\vec{k}_{\perp}\cdot\vec{\theta}\chi} P_{\Phi}[k_{\perp},z(\chi)] \label{App:vector2}\\
\langle \zeta_{a,\perp}^i\zeta_{b,\perp}^j\rangle&=&
4\int_0^{\textrm{min}\{\chi_{a}^0,\chi_{b}^0 \}} d\chi \, (\chi_{a}^0-\chi)(\chi_{b}^0-\chi)\nonumber\\
&\times&
\int\frac{d^2\vec{k}_{\perp}}{(2\pi)^2}\, k_{\perp}^i k_{\perp}^j \, e^{i\vec{k}_{\perp}\cdot\vec{\theta}\chi} P_{\Phi}[k_{\perp},z(\chi)],\qquad\label{App:tensor}
\end{eqnarray}
where in Eqs.~(\ref{App:vector2}, \ref{App:tensor}) $i,j=1,2$ denote the two components perpendicular to the line of sight.
Looking at the structure of Eqs.~(\ref{App:scalar}-\ref{App:tensor}) above
it is possible to notice that these quantities transform as a scalar, a vector and a symmetric tensor with respect to rotations in a plane perpendicular to the line of sight. We can therefore define right away the scalar $S_{ab}$ as
\begin{equation}
 S_{ab}\equiv\langle \zeta_{a,\parallel} \zeta_{b,\parallel} \rangle,\label{App:defSab}
\end{equation} 
and can simplify even more Eqs.~(\ref{App:vector2}, \ref{App:tensor}) by extracting the components of the vector and of the tensor. In the case of  the vector, for instance, it is possible to notice that the vector component is aligned along the displacement vector $\langle \zeta_{a,\perp}^j\zeta_{b,\parallel}\rangle\sim r_{\perp}^j$. We can then define
\begin{eqnarray}
 V_a&\equiv& \frac{r_j}{r_{\perp}}
\langle\zeta_{a,\perp}^j\zeta_{b,\parallel}\rangle\\
&=&4\int_0^{\textrm{min}\{\chi_{a}^0,\chi_{b}^0 \}} d\chi(\chi_a-\chi)
\nonumber\\
&\times&\int\frac{k^2\,dk}{2\pi}P_{\Phi}[k,z(\chi)]J_1(k\chi\theta)\label{App:vector}.
\end{eqnarray} 
Similarly it is possible to decompose the tensor part into its trace and its off diagonal traceless symmetric part (cfr. \cite{Lewis:2006fu}). Letting
\begin{equation}\label{tensdec}
\langle \zeta_{a,\perp}^i\zeta_{b,\perp}^j\rangle=T_{ab}\delta^{ij}_{\perp}-D_{ab} \hat{R}_{\perp}^{ij},
\end{equation}
where $\hat{R}_{\perp}^{ij}$ is the symmetric traceless tensor defined by
\begin{equation}
\hat{R}_{\perp}^{ij}\equiv\frac{1}{r_{\perp}^2}\left[r_{\perp}^i r_{\perp}^j-\frac{r_{\perp}^2}{2}\delta^{ij}\right].
\end{equation}
and then using
\begin{eqnarray}
\delta_{ij,\perp}\langle \zeta_{a,\perp}^i\zeta_{b,\perp}^j\rangle&=&2T_{ab},\\
r_{\perp,i} r_{\perp,j} \langle \zeta_{a,\perp}^i\zeta_{b,\perp}^j\rangle&=& r_{\perp}^2 \left(T_{ab}-\frac{D_{ab}}{2}\right),\label{App:rirjzz}
\end{eqnarray}
together with the Limber approximation, it is possible to get to the following expressions for $T_{ab}$ and $D_{ab}$ (cfr. \cite{Lewis:2006fu})
\begin{eqnarray}
T_{ab}&=& 4\int_0^{\textrm{min}\{\chi_{a}^0,\chi_{b}^0 \}} d\chi \, (\chi_{a}^0-\chi)(\chi_{b}^0-\chi)\nonumber\\
&\times&\int\frac{k^3dk}{4\pi}J_0(k\chi\theta)\,P_{\Phi}(k,\chi),\\
D_{ab}&=& 4\int_0^{\textrm{min}\{\chi_{a}^0,\chi_{b}^0 \}} d\chi \, (\chi_{a}^0-\chi)(\chi_{b}^0-\chi)\nonumber\\
&\times&\int\frac{k^3dk}{2\pi}J_2(k\chi\theta)\,P_{\Phi}(k,\chi),\label{App:TandD}
\end{eqnarray}
Taking the limit $\theta\rightarrow 0$ of Eqs.~(\ref{App:scalar}, \ref{App:vector}, \ref{App:TandD}) it is possible to obtain the equivalent of the above expressions for the single source case. In particular, given that $J_1(0)=J_2(0)=0$, it is straightforward to notice that in this case $V_a=0$ and $D_{aa}=0$. We can then forget about the labels for $D_{ab}$ and from here on simply identifying it with $D$.

Let's notice that up to this point only the Limber approximation has entered the above derivation. It is then possible to proceed further by defining
\begin{equation}
 T\equiv\frac{1}{2}\left(T_{aa}+T_{bb}-2T_{ab}\right).\label{App:T}
\end{equation} 
Defining the following shorthand notation for the sake of brevity,
\begin{eqnarray}
 I_0(\chi)&\equiv&\int\frac{dk}{\pi}\,k^3\,P_{\Phi}(k,\chi),\\
 I_J(\chi)&\equiv&\int\frac{dk}{\pi}\,k^3\,P_{\Phi}(k,\chi)\,J_0(k\chi\theta),
\end{eqnarray} 
we can then rewrite Eq.~(\ref{App:T}) above as
\begin{eqnarray}
 T&=&\frac{1}{2}\Bigg[\int_0^{\chi_a}d\chi(\chi_a-\chi)^2\,I_0(\chi)+
\int_0^{\chi_b}d\chi(\chi_b-\chi)^2\,I_0(\chi)\nonumber\\
&+&\int_0^{\chi_a}d\chi(\chi_b-\chi)(\chi_a-\chi)\,I_J(\chi)\Bigg]\nonumber\\
&=&\frac{1}{2}\Bigg[\int_0^{\chi_a}d\chi\left[(\chi_a-\chi)^2+(\chi_b-\chi)^2\right]I_0(\chi)\nonumber\\
&+&\int_{\chi_a}^{\chi_b}d\chi(\chi_b-\chi)^2I_0(\chi)\nonumber\\
&-&2\int_0^{\chi_a}d\chi(\chi_a-\chi)(\chi_b-\chi)I_J(\chi)\Bigg]\nonumber\\
&=&\frac{1}{2}\Bigg[\int_0^{\chi_a}d\chi(\chi_b-\chi_a)^2\,I_0(\chi)\nonumber\\
&+&\int_{\chi_a}^{\chi_b}d\chi(\chi_b-\chi)^2I_0(\chi)\nonumber\\
&-&2\int_0^{\chi_a}d\chi(\chi_a-\chi)(\chi_b-\chi)[I_J(\chi)-I_0(\chi)]\Bigg]\label{App:T1}
\end{eqnarray} 
where without loss of generality we assumed that $\chi_a<\chi_b$. Equation (\ref{App:T1}) above is exact. We can proceed further by noting that $\chi_b-\chi_a=r_{\parallel}$ and that the integration limits of the second integral extend over $r_{\parallel}$. We then have
\begin{eqnarray}
 &&\int_0^{\chi_a}d\chi(\chi_b-\chi_a)^2\,I_0(\chi)
+\int_{\chi_a}^{\chi_b}d\chi(\chi_b-\chi)^2I_0(\chi)\nonumber\\
 &=&r_{\parallel}^2\left[\int_0^{\chi_a}d\chi\,I_0(\chi)+\int_{\chi_a}^{\chi_b}d\chi\,I_0(\chi)\frac{(\chi-\chi_b)^2}{r_{\parallel}^2}\right]\nonumber\\
&\approx&r_{\parallel}^2\int_0^{\chi_a}d\chi\,\int\frac{dk}{\pi}\,k^3\,P_{\Phi}(k,\chi)= 2 T_1(\chi_a,r_{\parallel}),
\end{eqnarray}  
where to get to the last line we used the fact that while the first integral scales as $r_{\parallel}^2\chi_a$, the second one scales only as $r_{\parallel}$ and it therefore provides a subdominant contribution. Finally, the last term of Eq.~(\ref{App:T1}) gives $T_2(\chi_a,r_{\perp})$ provided that we make the approximation $(\chi-\chi_b)\approx(\chi-\chi_a)$. Then
\begin{eqnarray}
 &-&\int_0^{\chi_a}d\chi(\chi_a-\chi)(\chi_b-\chi)[I_J(\chi)-I_0(\chi)]\nonumber\\
&\approx&\int_0^{\chi_a}d\chi(\chi_a-\chi)^2\int\frac{dk}{\pi}\,k^3\,P_{\Phi}(k,\chi)\,\left[1-J_0(k\chi\theta)\right]\nonumber\\
&=& T_2(\chi_a,r_{\perp}).
\end{eqnarray} 
Putting all the pieces together we have then shown that
\begin{equation}
 T=T_1(\chi_a,r_{\parallel})+T_2(\chi_a,r_{\perp}).
\end{equation} 
Finally to recover the matrix decomposition of the $Z^{ij}$ correlator we can appeal to the simmetry of the lensing displacements $\zeta$. Notice in fact that from Eqs.~(\ref{App:displacement_perp}, \ref{App:displacement_par}) the displacements are clearly invariant with respect to a rotation around the line of sight. We can then arbitrarily pick the coordinate system so that the sources lie in the $x-z$ plane, with the $z$ axis directed along line of sight to $\vec{x}_a$. The coordinates of the sources are then $\vec{x}_a=(0,0,\chi_a)$ and 
$\vec{x}_b=(r_{\perp},0,\chi_a+r_{\parallel})$ respectively, and since the displacement vector $\vec{r}$ makes an angle $\alpha$ with the line of sight we have that $r_{\perp}=r\,\sin(\alpha)$ and $r_{\parallel}=r\,\cos(\alpha)$. From here it is then straightforward to obtain the expression for $Z^{ij}$ as in the second line of Eq.~(\ref{Zdef}).

\subsection{Derivation of Eq.~(\ref{GenRes:LH})}

Finally notice that the correction term appearing in Eq.~(\ref{GenRes:Corrfunctexpansion}) due to lensing can be recast in the form
\begin{eqnarray}
 \langle AB\rangle_2
 &\approx&\frac{Z^{ij}}{r^2} 
\left[ r_i r_j \frac{d^2 \langle AB\rangle}{dr^2}+ r\frac{d \langle AB\rangle}{dr}\left(\delta_{ij}-\frac{r_i r_j}{r^2}\right)\right]\nonumber\\
&=&\frac{Z^{ij}}{r^2}\Big[r_i\,r_j\,r\frac{d}{dr}\left(\frac{1}{r}\frac{d\langle AB\rangle}{dr}\right)
\nonumber\\
&+&r\frac{d\langle AB\rangle}{dr}\delta^{ij}\Big].
\end{eqnarray} 
Then, using Eq.~(\ref{Zdef}) for $Z^{ij}$, which is valid with this choice of coordinates, we get
\begin{eqnarray}
 \langle AB\rangle_2&=&r\frac{d}{dr}\left(\frac{1}{r}\frac{d\langle AB\rangle}{dr}\right)\nonumber\\
&\times&\left[\left(T-\frac{D}{2}\right)\frac{r_\perp^2}{r^2}
-\left(V_a+V_b\right)\frac{r_{\perp}r_{\parallel}}{r^2}+S\frac{r_{\parallel}^2}{r^2}\right]\nonumber\\
&+&\frac{1}{r}\frac{d\langle AB\rangle}{dr}\left(2T+S\right),
\end{eqnarray} 
which reduces to Eq.~(\ref{GenRes:LH}) once we notice that $T_2$ and $D$ provide the dominant contributions.

\section{The exact but more restrictive treatement}\label{App:Exact}

Following the approach of \cite{Lewis:2006fu}, it is possible to obtain an ``exact'' expression for the exponential appearing in Eq.~(\ref{GenRes:expcorrfunc}) under the restrictive condition that the quantity $k_i(\zeta_a^i-\zeta_b^i)$ is gaussian distributed. 
If this condition holds then we can use the fact that
\begin{equation}
 \langle e^{iy}\rangle=e^{-\langle y^2\rangle/2},
\end{equation} 
to rewrite Eq.~(\ref{GenRes:expcorrfunc}) as
\begin{eqnarray}
\langle\tilde{A}\tilde{B}\rangle
&=& \int\frac{d^3k}{(2\pi)^3}
e^{i\vec{k}\cdot(\vec{x}_a-\vec{x}_b)}P_{AB}(k, \eta_a, \eta_b)\nonumber\\
&\times&
e^{-\frac{1}{2}\langle\left[\vec{k}\cdot(\vec{\zeta}_a-\vec{\zeta}_b)\right]^2\rangle}.\label{App:nonperturbative1}
\end{eqnarray}
We therefore need to calculate the value of the correlator
\begin{eqnarray}
\left\langle\left[\vec{k}\cdot(\vec{\zeta}_a-\vec{\zeta}_b)\right]^2\right\rangle&=& k_i k_j \Big[ \langle\zeta_a^i\zeta_a^j\rangle + \langle\zeta_b^i\zeta_b^j\rangle-\langle\zeta_a^i\zeta_b^j\rangle \nonumber\\ 
&-&\langle\zeta_b^i\zeta_a^j\rangle \Big].
\end{eqnarray}
It is possible to proceed further by taking into account the fact that the lensing-induced displacements are characterized by different expressions that depend on the direction of the displacement (whether it is parallel or perpendicular to the line of sight). This fact then suggests automatically the adoption of a cylindrical coordinate system for $k$ space. Then, considering for simplicity the case of $k_i k_j\langle\zeta_a^i\zeta_b^j\rangle$ we have
\begin{eqnarray}
 k_i k_j\langle\zeta_a^i\zeta_b^j\rangle 
= k_{\parallel}^2\langle\zeta_{a,\parallel}\zeta_{b,\parallel}\rangle+
k_{\parallel}k_{i,\perp}\langle\zeta_{a,\parallel}\zeta_{b,\perp}^i\rangle
&&\nonumber\\
+
k_{j,\perp}k_{\parallel}\langle\zeta_{a,\perp}^j\zeta_{b,\parallel}\rangle+k_{i,\perp}k_{j,\perp}\langle\zeta_{a,\perp}^i\zeta_{b,\perp}^j\rangle.&&
\end{eqnarray} 
Finally, letting $k_{i,\perp}r_{\perp}^i=k_{\perp}r_{\perp}\cos(\gamma)$
and using Eqs.~(\ref{App:defSab}-\ref{App:rirjzz}) it is possible to recast the different terms appearing in the above equation as
\begin{eqnarray}
k_{\parallel}^2\langle\zeta_{a,\parallel}\zeta_{b,\parallel}\rangle&=&
k_{\parallel}^2 S_{ab},\\
k_{\parallel}k_{i,\perp}
\langle\zeta_{a,\parallel}\zeta_{b,\perp}^i\rangle
&=&-k_{\parallel}k_{\perp}\cos(\gamma)V_b,\\
k_{i,\perp}k_{j,\perp}\langle\zeta_{a,\perp}^i\zeta_{b,\perp}^j\rangle&=&
k_{\perp}^2\left[T_{ab}-\frac{D_{ab}}{2}\cos(2\gamma) \right].
\end{eqnarray} 
Equation (\ref{App:nonperturbative1}) above then becomes
\begin{eqnarray}
 \langle\tilde{A}\tilde{B}\rangle
&=& \int\frac{k_{\perp}dk_{\perp}dk_{\parallel}d\gamma}{(2\pi)^3}
e^{i\left[k_{\parallel}r_{\parallel}+k_{\perp}r_{\perp}\cos(\gamma)\right]}P_{AB}(k, \eta_a, \eta_b)\nonumber\\
&\times&
\exp\Bigg\{
k_{\parallel}^2\left( S_{ab}-\frac{S_{aa}+S_{bb}}{2}\right)\nonumber\\
&+&k_{\parallel}k_{\perp}\cos(\gamma)\left(V_a-V_b\right)\nonumber\\
&+& k_{\perp}^2\left[T_{ab}-\frac{D_{ab}}{2}\cos(2\gamma)-\frac{T_{aa}+T_{bb}}{2}
\right]
\Bigg\},\label{App:nonperturbative2}
\end{eqnarray} 
where for sake of clarity we have written down explicitly the measure of the integral and we should remind that since the displacement vector $\vec{r}$ makes an angle $\alpha$ with the line of sight $r_{\parallel}=r\cos(\alpha)$ and $r_{\perp}=r\sin(\alpha)$.

If we now consider the ``purely angular limit'' in which only displacements that are perpendicular to the line of sight are taken into account (as in \cite{Vallinotto:2007mf}) we restrict ourselves to the case in which the scalar and vector terms of the correlators vanish (the vector term because $V_a=V_b$, the scalar term because $S_{ab}\simeq S_{aa}=S_{bb}$). In this case Eq.~(\ref{App:nonperturbative2}) then simplifies considerably into
\begin{eqnarray}
 \langle\tilde{A}\tilde{B}\rangle
&=& \int\frac{k_{\perp}dk_{\perp} d\gamma}{(2\pi)^2}
e^{i k_{\perp}r_{\perp}\cos(\gamma)}P_{AB}(k, \eta)\nonumber\\
&\times&
\exp\Bigg\{
k_{\perp}^2\left[T_{ab}-\frac{D_{ab}}{2}\cos(2\gamma)-T_{aa}
\right]
\Bigg\}.\label{App:nonperturbative3}
\end{eqnarray} 
This expression agrees with the one obtained by Challinor and Lewis \cite{Lewis:2006fu} for the CMB. The point that need to be stressed, however, is that it both Eq.~(\ref{App:nonperturbative2}) and Eq.~(\ref{App:nonperturbative3}) are general expressions that are valid for \textit{any kind of sources}. Finally, the above integral can be carried out exactly by expanding $\exp(D_{ab}\cos(2\gamma)/2)$ in power series and then integrating term by term.

Finally, it seems necessary to point out here the difference between the two approaches and the assumptions that are underlying both of them. The ``non-perturbative'' approach requires $k_i(\zeta_a^i-\zeta_b^i)$ to be gaussian distributed. Once this price is paid, the apparent reward is to be able to take into account the full exponential, that is all the infinite terms appearing in its series expansion. On the other hand, the ``perturbative'' approach does not require such an assumption simply because higher order effects would contribute to higher order correlators and these are automatically discarded when the series expansion of the exponential is truncated to second order. It seems necessary to point out, however, that the increased accuracy that can be attained adopting the first approach is actually hard to evaluate. If non-linearities are present (and this is the case when considering that the lensing displacements are proportional to the gradient of the gravitational potential, which goes non-linear at late epochs/low redshift) it is then questionable whether summing all the terms appearing in the exponential would really lead to a \textit{consistently} more accurate result.

\end{document}